\documentclass[11pt]{article}
\usepackage{fancyhdr}
\usepackage{isomath}
\usepackage{amsmath}
\usepackage{amsbsy}
\usepackage{amssymb}
\usepackage{amscd}
\usepackage{amsfonts}
\usepackage{graphicx,color}
\usepackage{verbatim}
\usepackage{euscript}
\usepackage{alltt}
\usepackage{stmaryrd}
\usepackage{subfigure}
\usepackage{relsize}
\usepackage{enumitem}
\usepackage{url}

\DeclareGraphicsExtensions{.eps,.pdf}

\usepackage[margin=1in]{geometry}

\newcommand{\p}{\partial}
\newcommand{\dee}{\mathcal{D}}
\newcommand{\scl}{\mathcal{L}}

\date{}
\begin{document}
\title{Variational principles for nonlinear PDE systems via duality}

\author{Amit Acharya\thanks{Department of Civil \& Environmental Engineering, and Center for Nonlinear Analysis, Carnegie Mellon University, Pittsburgh, PA 15213, email: acharyaamit@cmu.edu.}}
\maketitle

\begin{abstract}
\noindent A formal methodology for developing variational principles corresponding to a  given nonlinear PDE system is discussed. The scheme is demonstrated in the context of the incompressible Navier-Stokes equations, systems of first-order conservation laws, and systems of Hamilton-Jacobi equations.
\noindent 
\end{abstract}

\section{Introduction}\label{sec:intro}
This work answers a question raised in \cite{acharya2021action}, namely, identifying the basic ingredients necessary for developing a variational principle all, or some, of whose Euler-Lagrange (E-L) equations are a given system of PDE; the functional to be developed is required to have space-time derivatives of its fundamental fields in `more-than-linear' combinations. Such a question arose from the purely practical issue of developing a basis for application of Effective Field Theory techniques in Physics (cf.~\cite{kleinert1989gauge_1, beekman2017dual1, beekman2017dual}) to the system of nonlinear dislocation dynamics \cite{acharya2021action} in continuum mechanics and materials science. Despite the success in formulating an appropriate action functional, that effort also exposed a certain flexibility in the adopted scheme whose details remained to be understood. Here, we are able to understand those details and abstract out the essence of the technique. The idea is then demonstrated on a wider setting of important classes of physical systems of nonlinear PDE. We note here that the question of finding a variational principle(s) (some of) whose E-L equations are a given system of PDE is different from the one adopted in the `Least-Squares Method,' (cf.~\cite[Ch.~10]{armstrong2019quantitative}), as explained in  \cite[Sec.~1]{seliger1968variational} and \cite{visintin}; the E-L equations of the Least Squares functional are not necessarily the PDE system from which the Least-Squares functional is developed. With the null minimizer requirement, minimizers of the Least Squares functional are solutions of the PDE system 
involved\footnote{As a related example, it is also instructive to consider the following: for $F:\mathbb{R}^2 \to \mathbb{R}$ let $F(u,\p_x u) = 0$ be a PDE for $u: (0,1) \to \mathbb{R}$. Let $H: \mathbb{R} \to \mathbb{R}$ be a function with an absolute minimum attained at $0$, and consider the functional $S[u] := \int_{(0,1)} dx \, H(F(u, \p_x u))$. Clearly any solution of the PDE is a minimizer of $S$. However, the Euler-Lagrange equation of $S$ is $\p_F H \, \p_u F - \p_x(\p_F H \, \p_{(\p_x u)} F) = 0$, which is of course not the same as the original PDE in any sense. I thank Gautam Iyer for a discussion of this case.}.
In the approach adopted herein, a \textit{family} of functionals is developed which satisfy the stated requirement. Mathematically rigorous considerations of the Least-Squares approach rely strongly on convex duality - the present approach relies on elements of convex duality even at a formal level.

This paper is organized as follows: in the following paragraph of this section we motivate the main idea of this work through an easy computation. In Secs.~\ref{sec:main_action}-\ref{sec:primal_mixed_action} we still use the `ad-hoc' procedure adopted in \cite{acharya2021action} to demonstrate our ideas in the context of the incompressible Navier-Stokes (N-S) equations for a homogeneous fluid. Related work can be found in \cite{kerswell1999variational, liu2007dual} which review earlier work in the physics and mechanics literature, including that of mathematicians C. Doering and P. Constantin; a sampling of the  mathematical literature on the matter can be found in \cite{ghoussoub2009anti,ortiz2018variational} and references provided therein (an important disclaimer here is that I am in no way an expert in the theory or practice of the N-S equations). In Sec.~6, the scheme is generalized to understand the fundamental restrictions, at least at a formal level, required to make it work, exploiting its full level of flexibility; this is  demonstrated in the context of a first-order system of conservation laws in divergence form and a system of Hamilton-Jacobi equations. Some closing observations are recorded in Sec.~\ref{sec:concl}. A word on notation: the Einstein summation convention for repeated indices is always invoked; $\Omega$ will represent a fixed domain in ambient, 3-d Euclidean space, and $[0,T]$ an interval of time.

Before moving on to the physical models we wish to discuss, we discuss the `toy' model of the heat equation in one space dimension and time to give the pattern of the typical computations that are involved in each case that follows, only nonlinear and in more space dimensions. Consider the heat equation
\[
\p_t \theta = \p_x \left(k \p_x \theta \right)
\]
and the functional
\begin{equation*}
\begin{aligned}
    \widehat{S}[\theta, \lambda] & :=  \int_{I \times [0,T]} dt dx \ \lambda ( \p_t \theta - \p_x (k \p_x \theta)) + H (\theta)\\
    & \  = \int_{I \times [0,T]} dt dx \ \theta (-\p_t \lambda - \p_x (k \p_x \lambda)) + H(\theta),
\end{aligned}
\end{equation*}
for \textit{any} convex function $H$ so that $p = H'(\theta)$ is uniquely solvable for $\theta(p)$, and assuming that $\lambda$ vanishes on the boundary of the interval $I \subset \mathbb{R}$ (representing the spatial domain).  The goal is to propose an `action' functional whose E-L equation is the original (system of) PDE in question, here the heat equation. Here, we have imposed the PDE (system) with a Lagrange multiplier field (to generate a scalar), exposed linear terms in the basic primal field(s) (here $\theta$) and added a convex term in the basic field, the latter two actions in anticipation of performing a Legendre transform. We note that the Lagrange multiplier field necessarily enters in a `linear' manner in the functional, even when the PDE is nonlinear (when the PDE contains nonlinear terms - as in the models considered later - all such nonlinear terms are combined additively with the function $H$ to define a function $M$). Motivated by the structure of $\widehat{S}$, now define
\begin{equation*}
\begin{aligned}
 M^*(p) & := p\, \theta(p) - H(\theta(p))\\
 p & := - ( - \p_t \lambda - \p_x (k \p_x \lambda) ).
\end{aligned}
\end{equation*}
With the above definitions, define the functional on a reduced state space
\[
S[\lambda] = \int_{I \times [0,T]} dt dx \ - M^*(p)
\]
whose first variation is given by
\[
\delta S = \int_{I \times [0,T]} dt dx \ - \p_p M^*(p) \delta p.
\]
Noting that $\p_p M^*(p) = \theta(p)$,
\[
\delta S = \int_{I \times [0,T]} dt dx \ \theta(p) \left( - \p_t \delta \lambda - \p_x \left (k \p_x \delta \lambda \right)\right)
\]
yielding the E-L equation
\[
\p_t (\theta(p)) - \p_x \left( k \p_x (\theta(p)) \right) = 0.
\]

With $H(\theta) = \frac{1}{2} \theta^2$, $M^*(p) = \frac{1}{2} p^2$ and $\theta(p) = p$, and the functional $S[\lambda]$ is bounded above. The Euler-Lagrange equation in terms of the function $\lambda$ is $\p^2_t \lambda - k^2 \p^4_x \lambda = 0$ which has the wrong sign when viewed as an initial value problem in time $t$ with dispersion relationship $\omega^2 = -k^2 m^4$ and eigenmode with wave number $m$ associated with growth factors $e^{\pm |k|m^2t}$. However, the functional can be maximized or its critical points approached by a gradient ascent in a `fake time' variable, say $s$, given by $\p_s \lambda = \frac{\delta S}{\delta \lambda}$, and it can be expected that the decaying solution (in $t$) is automatically picked up by such methods. For $k >0$, it can be checked that initial and boundary conditions on $\theta$ can be translated, non-uniquely, to constraints on the field $\lambda$ at the domain space-time boundaries. One also would seem to have the formal guarantee that any solution in the dual variable $\lambda$ that respects the boundary and initial constraints obtained from the primal problem must generate the unique solution to the primal problem through the mapping $\theta(x,t) = p(x,t) = \p_t \lambda(x,t) + k \,\p^2_x \lambda(x,t)$ (the mapping varies, of course, depending on the choice of the convex function $H$, but this conclusion remains unaltered).

We note that it is not our intent with the remarks above to suggest that the heat equation be solved in this manner, but only to motivate how the variational principles we develop henceforth may be `solved' or approximated. 

After completing this work, we became aware of the work of Brenier \cite{brenier2018initial}. There are strong connections of the ideas presented in this paper with those in \cite{brenier2018initial} - for instance, the functional \eqref{eq:dual_action} with $\hat{\nu} = 0$, $c = 1$, and $G(x) = \frac{1}{2}d x^2, d \to 0_+$ so that $G^*(\xi) = \frac{1}{2d} \xi^2$ formally recovers Brenier's variational principle for the Euler equation (up to accounting for the initial condition). Brenier does rigorous analysis to show existence of generalized solutions to his maximization principle which may be considered as encouraging for the ideas presented herein (the mathematical analysis in \cite{brenier2018initial} is beyond the scope of this paper).

\section{A `dual' action for the incompressible Navier-Stokes equation}\label{sec:main_action}

For the fields
\begin{equation}\nonumber
    \begin{aligned}
    \gamma &: \Omega \times [0,T] \to \mathbb{R}\\
    \lambda &: \Omega \times [0,T] \to \mathbb{R}^3
    \end{aligned}
\end{equation}
consider the functional
\begin{equation}\label{eq:dual_action}
    \begin{aligned}
    S_d[ \gamma, \lambda] = \int_{\Omega \times [0,T]} dt d^3x - \frac{1}{2} p_i \mathbb{K}_{ij} p_j - G^*(\xi),
    \end{aligned}
\end{equation}
where
\begin{equation}\label{eq:defs_d}
\begin{aligned}
p_k & := - [\hat{\nu} (\p_j\p_j \lambda_k + \p_k \p_i \lambda_i) - \p_k \gamma + \p_t \lambda_k]\\
\xi & := - \p_i \lambda_i\\
\mathbb{L}_{ij} (\nabla \lambda) & : = c \,\delta_{ij} + \p_j \lambda_i + \p_i \lambda_j; \qquad \qquad \mathbb{K}_{ij} (\nabla \lambda) := (\mathbb{L}(\nabla \lambda))^{-1}_{ij}; \qquad \qquad \mathbb{K}_{ij} = \mathbb{K}_{ji}\\
\omega(\xi) & := G'{}^{-1} (\xi)\\
G^*(\xi) & :=  \omega(\xi) \xi - G(\omega(\xi))
\end{aligned}
\end{equation}
for $c > 0$ an \textit{arbitrary}, non-dimensional constant, $G:\mathbb{R} \to \mathbb{R}$ an \textit{arbitrary convex} (smooth) function, and $G':\mathbb{R} \to \mathbb{R}$ refers to the derivative of the function $G$ whose inverse function exists. 
\textit{We make the assumption here that $\mathbb{L}$ is an invertible matrix for all possible values of its argument} (presumably this can be arranged by taking a sufficiently large value of the constant $c$). For $\nu > 0$ the shear viscosity and $\rho_0 > 0$ the constant density of the homogeneous, incompressible fluid, $\hat{\nu} : = \frac{\nu}{\rho_0}$.

In what follows, we will not always explicitly write the arguments of $\mathbb{L}, \mathbb{K}, \omega$, and $G^*$.
 
The first variation of $S_d$, assuming all variations vanish on the boundary of $\Omega \times [0,T]$\footnote{Here, we are interested in interior field equations; natural `boundary' conditions can be inferred by not assuming the variations to vanish on the boundary of the space-time domain and appropriately utilizing applied initial and spatial boundary conditions of the primal problem.} is given by
\begin{equation}\label{eq:first_var}
\begin{aligned}
\delta S_d = \int_{\Omega \times [0,T]} dt d^3x \ & - p_i \mathbb{K}_{ik} [-(\hat{\nu}(\p_j \p_j \delta \lambda_k + \p_k \p_i \delta \lambda_i) - \p_k \delta \gamma + \p_t \delta \lambda_k)] \\
& - G^*{}'[-\p_i \delta \lambda_i]\\
& - p_i p_k \,\delta \mathbb{K}_{ik},
\end{aligned}
\end{equation}
and noting that 
\begin{equation*}
\delta \mathbb{K}_{ik} = - \mathbb{K}_{ij} \delta \mathbb{L}_{jm} \mathbb{K}_{mk} = - \mathbb{K}_{ij} \mathbb{K}_{mk} [ \p_j \delta \lambda_m + \p_m \delta \lambda_j]
\end{equation*}
along with the definition
\begin{equation}\label{eq:eq:vel_d}
v_k(p, \nabla \lambda) := \mathbb{K}_{ik}(\nabla \lambda) p_i
\end{equation}
we have
\begin{equation}\nonumber
\begin{aligned}
\delta S_d = \int_{\Omega \times [0,T]} dt d^3x \ & \hat{\nu} ( \delta \lambda_k \p_j \p_j v_k + \delta \lambda_i \p_i \p_j v_j )\\
 & + (\p_k v_k) \delta \gamma - (\p_t v_k) \delta \lambda_k \\
 & - \frac{1}{2} [ \p_j(v_j v_k) + \p_m (v_k v_m)] \delta \lambda_k\\
 & - \p_i (G^*{}') \delta \lambda_i.
\end{aligned}
\end{equation}
Noting that
\[
G^*(\xi) = \omega(\xi) \xi - G(\omega(\xi)) \ \ \mbox{and} \ \ G'(\omega(\xi)) = \xi \Longrightarrow G^*{}'(\xi) = \omega(\xi)
\]
and defining
\begin{equation}\label{eq:pressure}
   \mathcal{P} := \rho_0 \, \omega 
\end{equation}
 the following Euler-Lagrange equations are obtained:
\begin{equation}\label{eq:E-L}
\begin{aligned}
\delta \lambda_k : \qquad & -\p_t v_k - \p_j (v_k v_j) + \p_j \left( \hat{\nu} (\p_j v_k + \p_k v_j) - \frac{\mathcal{P}}{\rho_0} \delta_{kj} \right) = 0\\
\delta \gamma : \qquad & \p_k v_k = 0.
\end{aligned}
\end{equation}
These are the Navier-Stokes system for an incompressible, homogeneous fluid if the fields $v, \mathcal{P}$ (that are defined in terms of the fields $\lambda$ and $\gamma$) are interpreted as the \textit{velocity} and the \textit{pressure} fields, respectively.
\section{The primal action for \eqref{eq:dual_action} and its reduced state space}\label{sec:primal_action}
We motivate how the functional \eqref{eq:dual_action} was arrived at. With all definitions and notation of the previous section enforced and in terms of the fields
\begin{equation}\nonumber
\begin{aligned}
v: \Omega \times [0,T] &\to \mathbb{R}^3\\
\gamma : \Omega \times [0,T] & \to \mathbb{R}\\
\lambda : \Omega \times [0,T] & \to \mathbb{R}^3 \\
\omega : \Omega \times [0,T] & \to \mathbb{R}
\end{aligned}
\end{equation}
consider the functional
\begin{equation}\nonumber
    \begin{aligned}
    \widehat{S}_d[v, \gamma, \lambda, \omega] = \int_{\Omega \times [0,T]} dt d^3x \ & c \frac{1}{2} v_i v_i + G(\omega)\\
    & + \gamma (\p_i v_i) \\
   &  + \lambda_i (\p_j (-\omega \delta_{ij} + \hat{\nu} (\p_j v_i + \p_i v_j)) - \p_t v_i - \p_j( v_i v_j)),
    \end{aligned}
\end{equation}
where the incompressible Navier-Stokes equations have been enforced via Lagrange multipliers.

Each term in the action density has physical dimensions of $\frac{Length^2}{Time^2}$; multiplying through by $\rho_0$ gives units of energy per unit volume.

The main idea is to invoke a Legendre transform based change of variables and then consider the variational principle in a \textit{reduced state space}. Assuming for the moment that the Lagrange multiplier fields vanish on the boundary, we have
\begin{equation}\nonumber
\begin{aligned}
\widehat{S}_d[v,\gamma, \lambda, \omega] = \int_{\Omega \times [0,T]} dt d^3x \ & c \frac{1}{2} v_i v_i  + v_i v_j \p_j \lambda_i + G(\omega)\\
    & + (\hat{\nu} (\p_j\p_j \lambda_i + \p_i \p_k \lambda_k) - \p_i \gamma + \p_t \lambda_i) v_i \\
   &  + \omega \p_i \lambda_i.
\end{aligned}
\end{equation}
Working with the definitions \eqref{eq:defs_d} and \eqref{eq:eq:vel_d}, we affect a reduction in the state space of $\widehat{S}$ to define
\begin{equation}\label{eq:red_action_mixed}
\begin{aligned}
S_d[\gamma, \lambda] = \int_{\Omega \times [0,T]} dt d^3x \ & \frac{1}{2} v_i(p, \nabla \lambda) \mathbb{L}_{ij}(\nabla \lambda) v_j(p, \nabla \lambda)  - p_i v_i(p, \nabla \lambda) + G(\omega(\xi))  - \omega(\xi) \xi.
\end{aligned}
\end{equation}
 Using the definitions \eqref{eq:defs_d} and \eqref{eq:eq:vel_d} once again and noting that
 \begin{equation}\label{eq:v-p}
 v_i p_i - \frac{1}{2} v_i \mathbb{L}_{ij} v_j = \frac{1}{2} p_i \mathbb{K}_{ik} p_k,
 \end{equation}
 we note that \eqref{eq:red_action_mixed} is the functional \eqref{eq:dual_action}.
\section{A `mixed' action for the incompressible Navier-Stokes equation}\label{sec:mixed_action}

For the fields
\begin{equation}\nonumber
    \begin{aligned}
    A &: \Omega \times [0,T] \to \mathbb{R}^{3 \times 3}_{sym}\\
    \gamma &: \Omega \times [0,T] \to \mathbb{R}\\
    \lambda &: \Omega \times [0,T] \to \mathbb{R}^3\\
    \omega &: \Omega \times [0,T] \to \mathbb{R}
    \end{aligned}
\end{equation}
consider the functional
\begin{equation}\label{eq:mixed_action_1}
    \begin{aligned}
    S_m[A, \gamma, \lambda, \omega] = \int_{\Omega \times [0,T]} dt d^3x - \frac{1}{2} p_i \mathbb{K}_{ij} p_j \mp R^*(\tau) + \omega \p_i \lambda_i,
    \end{aligned}
\end{equation}
where
\begin{equation}\label{eq:defs}
\begin{aligned}
p_k & := - [-\p_j A_{kj} - \p_k \gamma + \p_t \lambda_k]\\
\tau_{kl} & := \mp [- A_{kl} - \hat{\nu}  (\p_l \lambda_k + \p_k \lambda_l)]\\
\mathbb{L}_{ij} (\nabla \lambda) & : = c \, \delta_{ij} + \p_j \lambda_i + \p_i \lambda_j; \qquad \qquad \mathbb{K}_{ij}(\nabla \lambda) := (\mathbb{L}(\nabla \lambda))^{-1}_{ij}; \qquad \qquad \mathbb{K}_{ij} = \mathbb{K}_{ji}\\
D_{ij}(\tau) &:= \left(R'^{-1}\right)_{ij} (\tau)\\
R^*(\tau) &:= D_{ij}(\tau) \tau_{ij} - R(D(\tau))
\end{aligned}
\end{equation}
for $R:\mathbb{R}^{3 \times 3}_{sym} \to \mathbb{R}$ an arbitrary \textit{convex} function on the space of symmetric tensors, and $R'_{ij}:\mathbb{R}^{3 \times 3}_{sym} \to \mathbb{R}^{3 \times 3}_{sym}$ refers to the function $\p_{D_{ij}} R$ and $R'^{-1}_{ij}$ to its inverse function. 

The matrix field $\mathbb{L}$ and the constant $\hat{\nu}$ are defined exactly as in Sec. \ref{sec:main_action}.

The first variation of $S_m$, assuming all variations vanish on the boundary of $\Omega \times [0,T]$ (with the same understanding for what needs to be done to include natural boundary and initial conditions) is given by
\begin{equation}\label{eq:first_var_1}
\begin{aligned}
\delta S_m = \int_{\Omega \times [0,T]} dt d^3x \ & - p_i \mathbb{K}_{ik}(\nabla \lambda) [ - ( - \p_j \delta A_{kj} - \p_k \delta \gamma + \p_t \delta \lambda_k) ]\\
& \mp R^{*}{}'_{kl} [ \mp ( - \delta A_{kl} - 2\hat{\nu} \p_l \delta \lambda_k) ]\\
& + \delta \omega \p_i \lambda_i - (\p_i \omega) \delta \lambda_i - \frac{1}{2} p_i p_k \delta \mathbb{K}_{ik},
\end{aligned}
\end{equation}
and noting that 
\begin{equation*}
\delta \mathbb{K}_{ik} = - \mathbb{K}_{ij} \delta \mathbb{L}_{jm} \mathbb{K}_{mk}
\end{equation*}
we have
\begin{equation}\nonumber
\begin{aligned}
\delta S_m = \int_{\Omega \times [0,T]} dt d^3x \ & \p_j(p_i \mathbb{K}_{ik}(\nabla \lambda)) \delta A_{kj} + \p_k (p_i \mathbb{K}_{ik}(\nabla \lambda)) \delta \gamma - \p_t (p_i \mathbb{K}_{ik}(\nabla \lambda)) \delta \lambda_k \\
& - R^*_{kl} \delta A_{kl} + \p_l (2\hat{\nu} R^*{}'_{kl}) \delta \lambda_k + (\p_i \lambda_i) \delta \omega\\
& - \frac{1}{2} [ \delta \lambda_m \p_j (\mathbb{K}_{ij}(\nabla \lambda) p_i \mathbb{K}_{mk}(\nabla \lambda) p_k) + \delta \lambda_j \p_m(\mathbb{K}_{ij}(\nabla \lambda) p_i \mathbb{K}_{mk}(\nabla \lambda) p_k)],
\end{aligned}
\end{equation}
leading to the Euler-Lagrange equations
\begin{equation}\label{eq:E-L_1}
\begin{aligned}
\delta A_{ik} : & \qquad \frac{1}{2}  [ \p_j (p_i \mathbb{K}_{ik}(\nabla \lambda)) + \p_k (p_i \mathbb{K}_{ij}(\nabla \lambda))] - R^*{}'_{kj}(\tau) = 0 \\
\delta \gamma : & \qquad \p_k (p_i \mathbb{K}_{ik}(\nabla \lambda)) = 0 \\
\delta \lambda_k : & \qquad - \p_t ( p_i \mathbb{K}_{ik}(\nabla \lambda)) + \p_l (2 \hat{\nu} R^*{}'_{kl}(\tau) ) - \p_k \omega \\
& \qquad - \frac{1}{2} [ \p_j (\mathbb{K}_{ij}(\nabla \lambda) p_i \mathbb{K}_{km}(\nabla \lambda) p_m) + \p_m (\mathbb{K}_{ik}(\nabla \lambda) p_i \mathbb{K}_{mr}(\nabla \lambda) p_r) ]  = 0 \\
\delta \omega : & \qquad \p_i \lambda_i  = 0. 
\end{aligned}
\end{equation}

Defining a velocity and a pressure field as
\begin{equation}\label{eq:eq:vel}
v_k := \mathbb{K}_{ik} p_i; \qquad \qquad \mathcal{P} := \rho_0 \, \omega
\end{equation}
we note that the first three equations of \eqref{eq:E-L_1} imply the equations
\begin{equation}\label{eq:N-S}
\begin{aligned}
 \p_k v_k  & = 0 \\
- \p_t v_k - \p_j (v_k v_j) + \p_j \left[ \frac{\nu}{\rho_0} (\p_j v_k + \p_k v_j) - \frac{\mathcal{P}}{\rho_0} \delta_{kj} \right] & = 0,
\end{aligned}
\end{equation}
which is the Navier-Stokes system for a homogeneous, incompressible fluid.
\section{The primal action for \eqref{eq:mixed_action_1} and its reduced state space}\label{sec:primal_mixed_action}
We motivate how the functional \eqref{eq:mixed_action_1} was arrived at. With all definitions and notation of previous sections enforced and in terms of the fields
\begin{equation}\nonumber
\begin{aligned}
v: \Omega \times [0,T] &\to \mathbb{R}^3\\
D: \Omega \times [0,T] &\to \mathbb{R}^{3 \times 3}_{sym}
\end{aligned}
\end{equation}
consider the functional
\begin{equation}\nonumber
    \begin{aligned}
    \widehat{S}_m[v,D, \omega, A, \gamma, \lambda] = \int_{\Omega \times [0,T]} dt d^3x \ & c \frac{1}{2} v_i v_i \pm R(D)\\
    & + A_{ij} (\p_j v_i - D_{ij}) \\
   &  + \gamma (\p_i v_i) \\
   &  + \lambda_i (\p_j (- \omega \delta_{ij} + 2 \hat{\nu} D_{ij} ) - \p_t v_i - \p_j( v_i v_j)),
    \end{aligned}
\end{equation}
where the incompressible Navier-Stokes equations have been enforced via Lagrange multipliers.

As before, we invoke a Legendre transform based change of variables and then consider the variational principle in a reduced state space. Assuming for the moment that the Lagrange multiplier fields vanish on the boundary, we have
\begin{equation}\nonumber
\begin{aligned}
\widehat{S}_m[v,D, \omega, A, \gamma, \lambda] = \int_{\Omega \times [0,T]} dt d^3x \ & c \frac{1}{2} v_i v_i \pm R(D)\\
    & + (- \p_j A_{ij} - \p_i \gamma + \p_t \lambda_i) v_i \\
   &  + (- A_{kl} - \hat{\nu} (\p_l \lambda_k + \p_k \lambda_l)) D_{kl} \\
   &  + \omega \p_i \lambda_i + v_i v_j \p_j \lambda_i.
\end{aligned}
\end{equation}
Working with the definitions \eqref{eq:defs} and \eqref{eq:eq:vel}, we affect a reduction in the state space of $\widehat{S}_m$ to define
\begin{equation}\label{eq:red_action}
\begin{aligned}
S_m[A, \gamma, \lambda, \omega] = \int_{\Omega \times [0,T]} dt d^3x \ & \frac{1}{2} v_i(p, \nabla \lambda) \mathbb{L}_{ij}(\nabla \lambda) v_j(p, \nabla \lambda)  - p_i v_i(p, \nabla \lambda) \\
 & \pm R(D(\tau))  \mp \tau_{kl} D_{kl}(\tau) + \omega \p_i \lambda_i.
\end{aligned}
\end{equation}
 Using the definitions \eqref{eq:defs} and \eqref{eq:eq:vel} once again and noting \eqref{eq:v-p}, we note that \eqref{eq:red_action} is the functional \eqref{eq:mixed_action_1}.
\section{Generalizations}\label{sec:generalizations}
In the previous sections and in \cite{acharya2021action} the kinetic energy was chosen as the added potential associated with the velocity field in the primal actions. It is however clear that apart from a solvability condition, this potential should be amenable to a (more or less) arbitrary choice. In this section we demonstrate this feature of the proposed scheme through a discussion of important classes of PDE.

In the following subsections, we will repeatedly make use of an \textit{assumption} and its consequence, which we write out in detail before proceeding. It is essentially related to a Legendre transform in the presence of a parameter. For $M: \mathbb{R}^n \times \mathbb{R}^m \to \mathbb{R}$, $H: \mathbb{R}^n \to \mathbb{R}$, $L \in \mathbb{R}^m$, and $F:\mathbb{R}^n \to \mathbb{R}^m$ satisfying
\begin{equation}\label{eq:M_def}
M(U,L) = H(U) - L \cdot F(U),
\end{equation}
\textit{we assume that for given $P$ and $L$, there exists a unique function $U(P,L)$ which satisfies the relation}
\begin{equation}\label{eq:implicit_fn}
P = \p_U M(U(P,L),L).
\end{equation}
\textit{for all likely values of $L$ and $P$ to be encountered}, i.e., the algebraic system of equations $P = \p_U M(U, L)$ is uniquely solvable for $U$ in terms of $P, L$. If $M$ were to be convex in $U$ for all $L$, then such a condition would certainly hold.
In the following, the function $F$ will be specified from the physical PDE system to be solved and the function $H$ will be free to choose, so this condition is essentially a constraint on the choice of the class of functions to which $H$ belongs.

Assuming the above, consider
\begin{equation}
    \label{eq:M*_def}
    \begin{aligned}
        M^*(P,L) & := U(P,L)\cdot P - M(U(P,L), L)\\
        \Longrightarrow \p_L M^*(P,L) &=  \p_L U_i(P,L) P_i - \p_{U_i} M(U(P,L), L) \p_L U_i (P,L) - \p_L M(U(P,L), L) \\
        & = - \p_L M(U(P,L), L),
    \end{aligned}
\end{equation}
due to \eqref{eq:implicit_fn}. Given the form of \eqref{eq:M_def}, we have
\begin{equation}
    \label{eq:exp_deriv}
    \p_L M^*(P,L) = F(U(P,L)).
\end{equation}
We also have that
\begin{equation}
    \label{eq:dM*dp}
    \p_P M^*(P,L) =  \p_P  U_i(P,L) P_i + U(P,L) -  \p_{U_i} M (U(P,L),L) \p_P U_i (P, L) = U(P, L)
\end{equation}
by \eqref{eq:implicit_fn}.
\subsection{Dual variational principles for first-order systems of conservation laws in divergence form}
The PDE system of interest is of the form
\begin{equation}
    \label{eq:conserv_law}
    \p_t u_I + \p_i f_{Ii}(u) = 0,
\end{equation}
where $I = 1, \ldots ,n$, and $i = 1, \ldots, d$, $d$ being the number of space dimensions.

In terms of the field
\begin{equation*}
    \begin{aligned}
        \lambda &: \Omega \times [0,T] \to \mathbb{R}^n
    \end{aligned}
\end{equation*}
 consider the functional
\begin{equation}
    \label{eq:dual_conserv_law}
    S_{cl}[p, \lambda] = \int_{\Omega \times [0,T]} dt d^3x \, - M^*(p, \nabla \lambda),
\end{equation}
where $(\nabla \lambda)_{Ij} = \p_j \lambda_I$, along with the identifications $u = U$, $p = P$, $\nabla \lambda = L$, $m = nd$, $F = f$ in \eqref{eq:M_def}-\eqref{eq:implicit_fn} and the definitions
\begin{equation}
    \label{eq:conserv_law_defs}
    \begin{aligned}
    p_I & := \p_t \lambda_I \\
    M(u(p, \nabla \lambda), \nabla \lambda) & := H(u(p, \nabla \lambda)) - \nabla \lambda \cdot f(u(p, \nabla \lambda))\\
    M^*(p, \nabla \lambda) &:= u(p, \nabla \lambda) \cdot p - M(u(p, \nabla \lambda), \nabla \lambda)
    \end{aligned}
\end{equation}
for \textit{any} choice of the function $H$ that allows \eqref{eq:implicit_fn} to hold. Then the first variation is given by 
\begin{equation*}
    \delta S_{cl} = \int_{\Omega \times [0,T]} dt d^3x \, - \left( \p_p M^* \cdot \p_t \delta \lambda + \p_{\,\nabla \lambda} M^* \cdot \nabla \delta \lambda \right)
\end{equation*}
which leads to the E-L equations
\begin{equation*}
    \p_t u_I(p,\nabla \lambda) + \p_i f_{Ii} (u(p, \nabla \lambda)) = 0
\end{equation*}
by \eqref{eq:exp_deriv}-\eqref{eq:dM*dp}. Thus, every solution of the E-L equations of \eqref{eq:dual_conserv_law} defines a solution of \eqref{eq:conserv_law} through the definition of the function $U(P,L)$ related to \eqref{eq:implicit_fn}.

We also note that for given $f$, each member of the entire class of functions $M^*$, defined through a choice of an admissible state function $H$,  satisfies the conservation law \eqref{eq:conserv_law} in the sense
\begin{equation}\label{eq:entropies}
    \p_t \left(\p_p M^* \right)_I + \p_i \left( \p_{\,\nabla \lambda} M^* \right)_{Ii} = 0.
\end{equation}
\subsection{Dual variational principle for a second order system of Hamilton-Jacobi equations}
Consider the system of Hamilton-Jacobi equations
\begin{equation}
\label{eq:H-J_system}
\begin{aligned}
    \p_t u_I & = f_I \left( u, B, C \right)\\
    \p_i u_I & = B_{Ii}\\
    \p_i \p_j u_I & = C_{Iji}
    \end{aligned}
\end{equation}
where $f_I$ is a smooth function of its arguments.

In terms of the fields
\begin{equation*}
    \begin{aligned}
    \lambda &: \Omega \times [0,T] \to \mathbb{R}^n\\
    \gamma &: \Omega \times [0,T] \to \mathbb{R}^{n \times d}\\
    \rho &: \Omega \times [0,T] \to \mathbb{R}^{n \times d \times d}
    \end{aligned}
\end{equation*}
and the definitions
\begin{equation}
    \label{eq:H-J_P_def}
    \begin{aligned}
    P & := \left( \p_t \lambda + \nabla \cdot \gamma - \nabla^2 : \rho, \gamma, \rho \right)\\
    L & := \lambda\\
    F & := f, \\
    \end{aligned}
\end{equation}
where $(\nabla \cdot \gamma)_I = \p_i \gamma_{Ii}$ and $(\nabla^2 : \rho)_I = \p_j \p_i \rho_{Iij}$, we consider functions $H:\mathbb{R}^n \times \mathbb{R}^{n \times d} \times \mathbb{R}^{n \times d \times d} \to \mathbb{R}$ such that, for the generic element in its domain referred to as 
\[
U := ( u, B, C),
\]
the function $M$ in \eqref{eq:M_def} is defined with the solvability property \eqref{eq:implicit_fn}, and in terms of it, the function $M^*$ in \eqref{eq:M*_def}. 

Consider the functional
\begin{equation}
    \label{eq:H-j_dual}
    S_{HJ} [\lambda, \gamma, \rho] = \int_{\Omega \times [0,T]} dt d^3x - M^*(P, L),
\end{equation}
whose E-L equations are (assuming $(\lambda, \gamma, \rho)$ have compact support on $\Omega$)
\begin{equation*}
    \begin{aligned}
    - \p_t u_I(P,L) + f_I(U(P,L)) & = 0 \\
    - \p_i u_I(P,L) + B_{Ii}(P, L) = 0 \\
    -  \p_i \p_j u_I(P,L) + C_{Iji} (P, L) = 0.
    \end{aligned}
\end{equation*}

We note that the static system $\p_i f_{Ii}(B) = 0$, $B_{Ii} = \p_i u_I$ can be dealt with as an Hamilton-Jacobi system as well as by taking account of its conservation structure by the proposed technique, with $L = \nabla \lambda$ in the latter case.
\section{Concluding remarks}\label{sec:concl}
The proposed scheme for generating variational principles for nonlinear PDE systems may be abstracted and summarized as follows: We first pose the given system of PDE as a \textit{first-order} system (introducing extra fields representing (higher-order) space and time derivatives of the fields of the given system); as before let us denote this collection of primal fields by $U$. `Multiplying' the primal equations by dual Lagrange multiplier fields, the collection denoted by $D$, adding a function $H(U)$, solely in the variables $U$ (the purpose of which, and associated requirements, will be clear shortly), and integrating by parts over the space-time domain, we form a `mixed' functional in the primal and dual fields given by
\begin{equation*}
    \widehat{S}_H [U,D] = \int_{[0,T]\times \Omega} dt d^3x \ \scl_H (\dee,U)
\end{equation*}
where $\dee$ is a collection of local objects in $D$ and at most its first order derivatives. We then require that the family of functions $H$ be such that it allows the definition of a function $U_H(\dee)$ such that
\begin{equation*}
    \frac{\p \scl_H}{\p U} (\dee, U_H(\dee)) = 0
\end{equation*}
so that the \emph{dual} functional, defined solely on the space of the dual fields $D$, given by
\begin{equation*}
    S_H[D] = \int_{[0,T]\times \Omega} dt d^3x \ \scl_H(\dee,U_H(\dee))
\end{equation*}
has the first variation
\begin{equation*}
    \delta S_H = \int_{[0,T]\times \Omega} dt d^3x \ \frac{\p \scl_H}{\p \dee} \delta \dee.
\end{equation*}
By the process of formation of the functional $\widehat{S}_H$, it can then be seen that the (formal) E-L equations arising from $\delta S_H$ have to be the original first-order primal system, with $U$ substituted by $U_H(\dee)$, regardless of the $H$ employed.

Thus, the proposed scheme may be summarized as follows: we wish to pursue the following (local-global) critical point problem
\begin{equation*}
   \begin{smallmatrix} \mbox{extremize}\\ D\end{smallmatrix} \int_{[0,T]\times \Omega} dt d^3x \ \begin{smallmatrix} \mbox{extremize}\\ U\end{smallmatrix} \  \scl_H (\dee(t,x),U),
\end{equation*}
where the pointwise extremization of $\scl_H$ over $U$, for fixed $\dee$, is made possible by the choice of $H$.

Furthermore, assume the Lagrangian $\scl_H$ can be expressed in the form
\begin{equation*}
    \scl_H(\dee, U) := - P(\dee)\cdot U + f(U,D) + H(U)
\end{equation*}
for some function $P$ defined by the structure of the primal first-order system ((linear terms in) first derivatives of $U$ after multiplication by the dual fields and integration by parts always produce such terms), and for some function $f$ which, when non-zero, does not contain any linear dependence in $U$. Our scheme requires the existence of a function $U_H$ defined from `solving $\frac{\p \scl}{\p U} (\dee, U) = 0$ for $U$,' i.e.~$\exists \  U_H(P(\dee),\dee)$ s.t. the equation
\begin{equation*}
    - P(\dee) + \frac{\p f}{\p U}(U_H(P(\dee),\dee), \dee) + \frac{\p H}{\p U}\left(U_H(P(\dee),\dee)\right) = 0
\end{equation*}
is satisfied. This requirement may be understood as follows: define
\begin{equation*}
    f(U, \dee) + H(U) =: M(U, \dee)
\end{equation*}
and assume that it is possible, through the choice of $H$, to make the function $\frac{\p M}{\p U}(U, \dee)$ \textit{monotone} in $U$ so that a function $U_H(P,\dee)$ can be defined that satisfies
\begin{equation*}
    \frac{\p M}{\p U}(U_H(P,\dee), \dee) = P, \quad \forall P.
\end{equation*}
Then the Lagrangian is
\begin{equation*}
    \scl(\dee, U_H(P(\dee),\dee)) = - P(\dee) \cdot U_H(P(\dee),\dee) + M(U_H(P(\dee),\dee), \dee) =: - M^*(P(\dee), \dee)
\end{equation*}
where $M^*(P,\dee)$ is the Legendre transform of the function $M$ w.r.t $U$, with $\dee$ considered as a parameter.

Thus, our scheme may also be interpreted as designing a concrete realization of abstract saddle point problems in optimization theory \cite{rockafellar1974conjugate}, where we exploit the fact that, in the context of `solving' PDE viewed as constraints implemented by Lagrange multipliers to generate an unconstrained problem, there is a good deal of freedom in choosing an objective function(al) to be minimized. We exploit this freedom in choosing the function $H$ to develop dual variational principles corresponding to general systems of PDE.

We conclude with a few observations and directions for future work: 
\begin{enumerate}
\item A detailed study of how boundary and initial conditions of the primal problem can be transferred to the same for the dual problem is warranted, most importantly for practical purposes of generating numerical approximations to the proposed formal mathematical scheme. 

As a first example, consider the dual formulation of the heat equation discussed in Sec.~\ref{sec:intro}. It is already clear that such a problem in space-time cannot be viewed as an initial value problem because of the lack of continuous dependence w.r.t initial data, as discussed. Thus, one has to consider the dual problem as a boundary-value problem in space-time. An important question to resolve then is whether doing so allows enough freedom for predicting the correct evolving, primal temperature profile. For this, it seems reasonable that the dual E-L equation ($\p^2_t \lambda - k^2 \p^4_x \lambda = 0$), admits two side conditions in the time ($t$) direction, and four conditions in the space ($x$) direction. For a unique evolution on the primal side, one has to apply the boundary conditions of the primal problem, say pure Dirichlet on temperature $\theta$, and an initial condition on $\theta$ as well, which translate to constraints on the fields of the dual problem through the mapping, $\theta(x,t) = \p_t \lambda(x,t) + k \,\p^2_x \lambda(x,t)$. Thus, for a solution of the dual problem in the space-time domain $(x_l, x_r) \times [0,T]$, specifying $\lambda(x, T)$ (arbitrarily) and $\theta(x, 0)$ along with $\lambda(x_l,t), \lambda(x_r,t)$ (arbitrarily), and $\theta(x_l, t), \theta(x_r, t)$ would seem to be a possibility, which should still allow enough freedom in the development of $\theta(x, T)$ through the mapping $\theta(x,T) = \p_t \lambda(x,T) + k \,\p^2_x \lambda(x,T)$. Such questions are the focus of ongoing work.

\item Setting $\hat{\nu} = 0$ in the functionals \eqref{eq:dual_action} and \eqref{eq:mixed_action_1} yields stationary principles for the incompressible Euler equations.

\item The mixed variational principle for the incompressible N-S equations in Secs.~\ref{sec:mixed_action}-\ref{sec:primal_mixed_action} affords the addition of added viscosity through the function $R$ in the solution of the dual problem - it is an interesting question whether this feature of the problem can help in the analysis and solution of problems of turbulence.

\item The pointwise invertibility of the matrix field $\mathbb{L}$ in Secs.~\ref{sec:main_action}-\ref{sec:primal_mixed_action} appears to be a key issue in the formalism. It would be interesting to understand the effect of the condition $\det(\mathbb{L}) = 0$ on the proposed scheme and whether there can be a relation between the Reynolds number and the value of $c$ for optimal performance of the scheme.

In the general setting, this condition translates to the validity of the assumption \eqref{eq:implicit_fn}.

\item Constrained by appropriate boundary and initial conditions, when the PDE system has unique solutions, it is clear that \textit{any} choice of the function $H$ within the admissible class, coupled with the mapping from the dual fields to the primal fields, must lead to the unique solution. In the absence of uniqueness, it is an interesting question whether specific choices of $H$ (along with the dual-to-primal mapping) act as a selection mechanism for picking up particular solutions. For instance, an appropriate quadratic choice of $H$ for the `simple' equation $\p_x f(B) = 0, B = \p_x u$ where $f$ is nonconvex results in an essentially semilinear second-order dual problem that is definitely simpler than the quasilinear primal problem. The structure of the dual problem in this case does not make the expectation of a smooth solution an absurd one. If this is indeed borne out in reality, then the solution to the primal problem that is defined through the dual-to-primal mapping may also be expected to be smooth.

\item In the context of first-order systems of conservation laws, what connection, if any, might exist between the large class of functions $M^*$ and entropies \cite{dafermos} of conservation laws is an important question to resolve.

\item There does not exist well-established computational approximation schemes (and for that matter, theory) for \emph{systems} of H-J equations. This work provides a variational structure for such systems which naturally lends itself to, say a finite element, discretization. Whether such an idea has any practical merit would be interesting to explore. If so, this can be very useful for certain systems related to the mechanics of fracture and plastic deformation of solids,  see \cite{morinaanalysis, zhang2015single, arora2020unification}.

\item For static conservation laws of the form $\p_i f_{Ii}(\nabla u) = 0, (\nabla u)_{Ij} := \p_j u_I$, the scheme produces variational principles even when there does not exist a potential $\psi(\nabla u)$ such that $f_{Ii} = \p_{(\nabla u)_{Ii}} \psi$.

\item The considerations herein show that even a given variational principle can be associated with a vastly different dual variational principle through the choice of the function $H$, by associating the former's E-L equations with the latter following the proposed scheme. This seems to open up fascinating points of convergence between apparently different classes of physical models described in terms of PDE and/or variational principles. A particularly intriguing question is whether the present considerations have any connection to the correspondences like AdS-CFT or AdS-CMT \cite{zaanen2015holographic, hartnoll2018holographic} in the string theory-high energy-condensed matter-gravitational physics nexus.

\item The proposed duality scheme seems to suggest that for a given PDE system, elements of the class of admissible potentials $H$ form symmetry operations for the system. The algebraic structure of this class of potentials, starting from whether they form a group, is of independent interest, as well as whether knowledge of such structure can help in the understanding of solutions to the PDE system.
\end{enumerate}
\section*{Acknowledgments}
This work was supported by the grant NSF OIA-DMR \#2021019. It is a pleasure to acknowledge discussions with Marshall Slemrod and Vladimir Sverak (VS) and to thank them for taking the time to take a look at the paper. I also thank VS for pointing me to \cite{armstrong2019quantitative} and for encouragement.

\bibliographystyle{IEEEtran}\bibliography{disloc_action_ref}
\end{document}